\DocumentMetadata{}
\documentclass[sigconf]{acmart}
\AtBeginDocument{%
  \providecommand\BibTeX{{%
    \normalfont B\kern-0.5em{\scshape i\kern-0.25em b}\kern-0.8em\TeX}}}

\copyrightyear{2025}
\acmYear{2025}
\setcopyright{acmlicensed}\acmConference[WSDM '25]{Proceedings of the
Eighteenth ACM International Conference on Web Search and Data
Mining}{March 10--14, 2025}{Hannover, Germany}
\acmBooktitle{Proceedings of the Eighteenth ACM International Conference on
Web Search and Data Mining (WSDM '25), March 10--14, 2025, Hannover,
Germany}
\acmDOI{10.1145/3701551.3703557}
\acmISBN{979-8-4007-1329-3/25/03}

\usepackage{subfigure}
\usepackage{enumitem}
\usepackage{multirow}
\usepackage{makecell}
\usepackage{algorithm}
\usepackage{algpseudocode}
\usepackage{amsmath}

\DeclareMathOperator*{\argmin}{arg\,min}
\usepackage{threeparttable}
\usepackage{booktabs}
\usepackage{dsfont}

\begin{document}

\title{Fusion Matters: Learning Fusion in Deep Click-through Rate Prediction Models}

\author{Kexin Zhang}
\authornote{Both authors contributed equally to this research. Work done when they were research interns at FiT, Tencent.}
\affiliation{
  \institution{Northwestern University}
  \city{Evanston, IL}
  \country{USA}
}
\email{kevin.kxzhang@gmail.com}
\orcid{0000-0003-2678-8556}

\author{Fuyuan Lyu}
\authornotemark[1]
\affiliation{
  \institution{McGill University \& MILA}
  \city{Montreal}
  \country{Canada}
}
\email{fuyuan.lyu@mail.mcgill.ca}
\orcid{0000-0001-9345-1828}

\author{Xing Tang}
\authornote{Co-corresponding Authors}
\affiliation{
  \institution{FiT, Tencent}
  \city{Shenzhen}
  \country{China}
}
\email{xing.tang@hotmail.com}
\orcid{0000-0003-4360-0754}

\author{Dugang Liu}
\authornotemark[2]
\affiliation{
  \institution{Shenzhen University}
  \city{Shenzhen}
  \country{China}
}
\email{dugang.ldg@gmail.com}
\orcid{0000-0003-3612-709X}

\author{Chen Ma}
\affiliation{
  \institution{City University of Hong Kong}
  \city{Hong Kong SAR}
  \country{China}
}
\email{chenma@cityu.edu.hk}
\orcid{0000-0001-7933-9813}

\author{Kaize Ding}
\affiliation{
  \institution{Northwestern University}
  \city{Evanston, IL}
  \country{USA}
}
\email{kaize.ding@northwestern.edu}
\orcid{0000-0001-6684-6752}

\author{Xiuqiang He}
\affiliation{
  \institution{FiT, Tencent}
  \city{Shenzhen}
  \country{China}
}
\email{xiuqianghe@tencent.com}
\orcid{0000-0002-4115-8205}

\author{Xue Liu}
\affiliation{
  \institution{McGill University}
  \city{Montreal}
  \country{Canada}
}
\email{xueliu@cs.mcgill.ca}
\orcid{0000-0001-5252-3442}

\renewcommand{\shortauthors}{Kexin Zhang et al.}

\begin{abstract}
The evolution of previous Click-Through Rate (CTR) models has mainly been driven by proposing complex components, whether shallow or deep, that are adept at modeling feature interactions. However, there has been less focus on improving fusion design. Instead, two naive solutions, stacked and parallel fusion, are commonly used. Both solutions rely on pre-determined fusion connections and fixed fusion operations. It has been repetitively observed that changes in fusion design may result in different performances, highlighting the critical role that fusion plays in CTR models. While there have been attempts to refine these basic fusion strategies, these efforts have often been constrained to specific settings or dependent on specific components. Neural architecture search has also been introduced to partially deal with fusion design, but it comes with limitations. The complexity of the search space can lead to inefficient and ineffective results. To bridge this gap, we introduce OptFusion, a method that automates the learning of fusion, encompassing both the connection learning and the operation selection. We have proposed a one-shot learning algorithm tackling these tasks concurrently. Our experiments are conducted over three large-scale datasets. Extensive experiments prove both the effectiveness and efficiency of OptFusion in improving CTR model performance. Our code implementation is available here\footnote{\url{https://github.com/kexin-kxzhang/OptFusion}}.
\end{abstract}



\maketitle

\section{Introduction}
\label{sec:intro}

Click-through rate (CTR) prediction is a vital task for commercial recommender systems and online advertising platforms, as it seeks to predict the likelihood that a user will click on a recommended item, such as a movie or advertisement~\cite{LR, ADS}.
As deep learning-based CTR models have advanced, researchers have developed various model architectures~\cite{FNN,IPNN,DeepFM,DCN,DCNv2,xDeepFM} to better capture feature interactions and enhance prediction performance.
These deep CTR models employ a combination of explicit and implicit components to represent feature interactions. Shallow components, including inner products~\cite{IPNN}, cross layer~\cite{DCN}, and Factorization Machines (FM)~\cite{FM,DeepFM}, are used to model these interactions explicitly. Concurrently, deep components like multi-layer perceptrons (MLP)~\cite{FNN} and Self-attention layer~\cite{Autoint} implicitly capture the complexity of feature interaction. 
With the improved deep and shallow components, deep CTR models can model feature interactions more effectively, leading to better prediction accuracy.

\begin{figure*}[!htbp]
    \centering
    \includegraphics[width=0.92\textwidth]{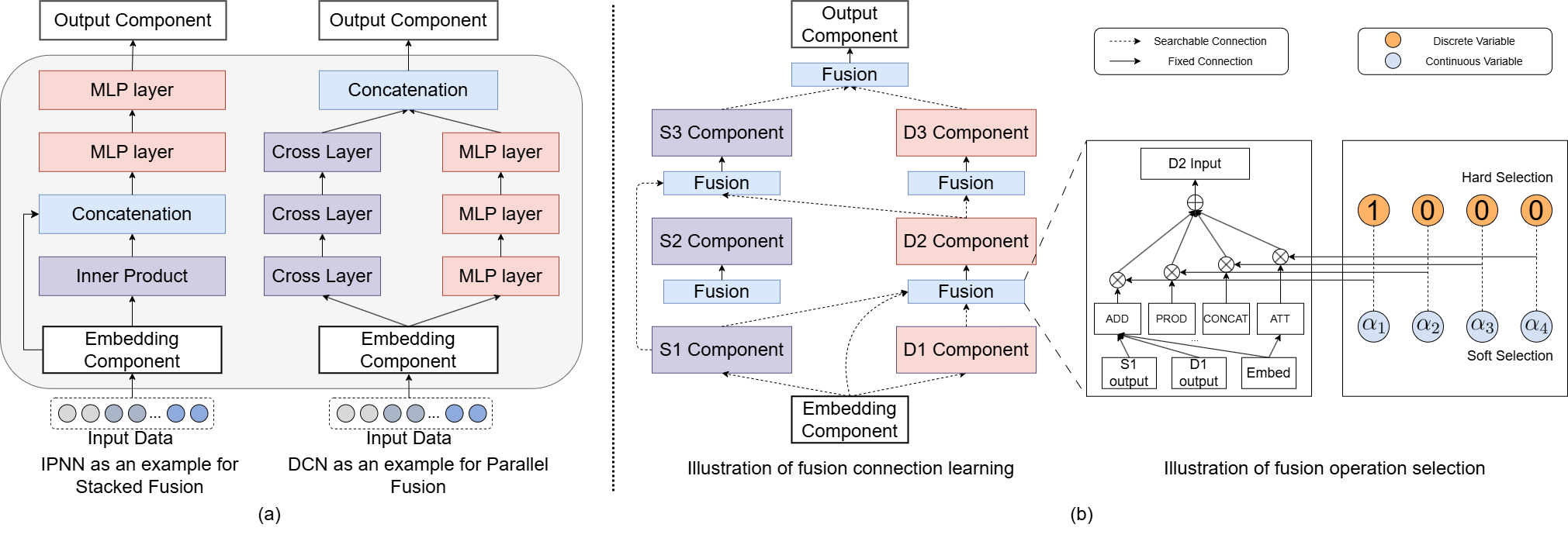}
    \vspace{-5pt}
    \caption{(a) Illustration figure about two mainstream fusion designs in deep CTR models. (b) The framework of OptFusion, which consists of one embedding component, $n$ cross components, $n$ deep components, and one output component as the candidate set of fusion connection search. 
    The selection of components is formed as a Directed Acyclic Graph (DAG). Fusion operation selection is also designed to fuse representations from lower-level components.}
    \label{fig:framework}
    \Description[The framework of OptFusion]{The framework of OptFusion consists of one embedding component, $n+1$ cross components, $n$ deep components, and one output component as the candidate set of fusion connection search. Fusion operation selection is also designed to fuse information extracted by lower-level components.}
\end{figure*}

Despite advances in CTR prediction through the enhancement of deep and shallow components, the design of fusion mechanisms has not been extensively studied. Fusion design is crucial as it involves aggregating representations from different model components. Previous works~\cite{DCN,IPNN}, as illustrated in Figure \ref{fig:framework}, have predominantly relied on two naive fusion designs: \textit{stacked} and \textit{parallel}.
In the stacked design, shallow components are typically placed before deep components and are trained sequentially~\cite{IPNN,FGCNN,FiBiNet}. For example, as depicted in Figure \ref{fig:framework} (a), IPNN~\cite{IPNN} uses the inner product as a shallow component, concatenating its output with original embeddings before feeding them into deep components.
Conversely, the parallel design involves the joint training of shallow and deep components.
Models like DCN~\cite{DCN} use concatenation for fusion,
while others, such as DeepFM~\cite{DeepFM} and Wide\&Deep~\cite{WDL}, employ addition to combine outputs from both components~\cite{WDL,xDeepFM,Autoint}.
In summary, these two naive fusion designs rely on \textbf{pre-defined fusion connections} and \textbf{fixed fusion operations} to fuse representations from both deep and shallow components. 
However, variations in fusion design can lead to substantial differences in performance across different datasets. For instance, DCNv2~\cite{DCNv2} with a parallel design may outperform its stacked counterpart on the MovieLens dataset while underperforming on the Criteo dataset. 
This inconsistent performance among different fusion designs across various datasets is also observed in MaskNet~\cite{MaskNet}.
These findings underscore the critical role of fusion in CTR predictions.

Attempts have been made to refine the above-mentioned fusion designs. 
EDCN~\cite{EDCN}, for instance, introduces a manually crafted complex fusion design that incorporates additional fusion connections and sophisticated fusion operations. Despite using the same components as its predecessor DCN~\cite{DCN}, EDCN achieves significant improvement in performance.
FinalMLP~\cite{FinalMLP} suggests the use of Multi-Head Bilinear Fusion Operations as a more effective means of integrating representations, outperforming naive fusion operations such as concatenation or addition. 
These examples demonstrate the potential of fusion design in significant improvements.
However, these proposed solutions~\cite{EDCN,FinalMLP} typically address the fusion learning problem under specific settings or depending on specific modules, failing to demonstrate the potential of fusion learning explicitly. 
A more general and adaptable approach to fusion learning remains a compelling challenge.

Other researchers have explored using neural architecture search (NAS) within CTR models to deal with fusion design challenges by defining a broader search space~\cite{AutoCTR,NASRec}. 
AutoCTR~\cite{AutoCTR} employs an evolutionary approach to simultaneously search for the optimal component type, raw feature input, fusion connections, and component-specific hyperparameters. However, this approach incurs a considerable training cost since each candidate architecture must be trained separately during the search phase.
NASRec~\cite{NASRec} takes a different approach by expanding the search space even further for better utility while leveraging weight-sharing techniques to mitigate the training cost during the search process. 
Despite these innovations, searching for fusion design in conjunction with other model structures, such as component types, within such a vast search space can make it much harder to obtain an optimal result.
In contrast, our research focuses solely on one of the critical aspects: fusion learning. By narrowing down the search space, better results and faster convergence speed can be achieved.
The issues mentioned above highlight the necessity for a comprehensive yet lightweight approach to fusion learning that simultaneously selects both fusion connections and operations.

To address these challenges, we introduce \textbf{OptFusion}, an automated fusion learning framework for deep CTR prediction. OptFusion aims to explore how fusion, both in terms of connections and operations, can impact CTR predictions and automatically identify the most suitable fusion design.
We propose a unified search space specifically tailored to the fusion learning process with shallow and deep components. This design allows each component to be connected to its predecessors. Such a design effectively focuses the search space on fusion learning and facilitates more efficient exploration compared to neural architecture search methods in CTR models.
Drawing inspiration from previous work~\cite{AssembleNet++}, we propose a one-shot learning algorithm that concurrently learns fusion connections and selects fusion operations. This algorithm considers the entanglement and mutual influence between fusion connections and operations, leading to better selection results.
Through rigorous evaluation on three widely-used benchmarks, we empirically demonstrate the effectiveness and efficiency of OptFusion. 
Additionally, our ablation studies have shown the orthogonality of OptFusion when combined with various components.
We summarize our contributions as follows:
\begin{itemize}[topsep=0pt,noitemsep,nolistsep,leftmargin=*]
    \item We investigate the importance of fusion, both connections and operations. A novel CTR fusion learning framework, namely OptFusion, is proposed. OptFusion can automatically learn suitable fusions in CTR models, containing both connection learning and operation selection.
    \item To better explore the proposed fusion space, a one-shot learning algorithm that simultaneously selects fusion connection and operation is proposed. Such a one-shot learning algorithm can better explore the search space due to entanglement between fusion operation and connection.
    \item We empirically evaluate OptFusion's efficiency and effectiveness on three large-scale datasets.
\end{itemize}


\section{Preliminary}
\label{sec:preliminary}

In this section, we first formulate the CTR prediction problem in Section \ref{sec:preliminary_ctr}. Then, we introduce the fusion learning for CTR prediction in Section \ref{sec:preliminary_fusion}. 


\subsection{CTR Prediction}
\label{sec:preliminary_ctr}
CTR prediction is a classic supervised binary classification problem~\cite{LR}. Given a dataset $\mathcal{D} = \{(\mathbf{x}, y)\}$ consisting of $N = |\mathcal{D}|$ instances with each containing a pair of user and item, the CTR prediction aims to predict whether the user would click the item. Here $\mathbf{x}$ denotes the input data instance, and $y \in \{0, 1\}$ denotes the label indicating whether the user clicked the item. 

In deep learning-based CTR prediction models, an embedding layer $\mathcal{E}$ is usually adopted to transform the input $\mathbf{x}$ with high-dimensional sparse raw features into low-dimensional dense embeddings $\mathbf{e}$. The embedding $\mathbf{e}$ can be obtained via embedding lookup~\cite{DeepFM}, formulated as $\mathbf{e} = \mathcal{E}(\mathbf{x})$. The embedding $\mathbf{e}$ is further fed into the CTR model, formulated as follows:
\begin{equation}
\label{eq:output_naive}
    \hat{y} = \mathcal{F} \left(\mathbf{e}, \Theta\right) = \mathcal{F} \left( \mathcal{E}(\mathbf{x}), \Theta \right)
\end{equation}
where $\hat{y}$ is the probability that a user will click a given item, $\mathcal{F}(\cdot)$ is the prediction model, and $\Theta$ is the corresponding trainable model parameters. Cross-entropy loss is commonly adopted to train $\Theta$, which can be formulated as follows:
\begin{equation}
\label{eq:loss_naive}
\begin{aligned}
\argmin_{\Theta} \mathcal{L}_{\mathcal{D}}(\Theta) = \sum_{(\mathbf{x}, y) \in \mathcal{D}} y\log{\hat{y}} + (1-y)\log{(1-\hat{y})}.
\end{aligned}
\end{equation}


\subsection{Fusion Learning in CTR}
\label{sec:preliminary_fusion}

In this section, we aim to take a deeper look at the CTR model $\mathcal{F}(\cdot)$. Here, we formulate the CTR model as an instance of the fusion design, written as:
\begin{equation}
    \mathcal{F} = \mathcal{P}(\mathcal{G} | \mathbf{c}, \mathbf{o}).
\end{equation}
Here $\mathcal{P}$ denotes the fusion learning, parameterized by fusion connection $\mathbf{c}$ and fusion operation $\mathbf{o}$. $\mathcal{G} = \{ \mathcal{G}^s, \mathcal{G}^d \}$ refer to the set of all components in the CTR model, with $\mathcal{G}^s$ and $\mathcal{G}^d$ represent the set of shallow and deep components, respectively. Various shallow components such as cross layer~\cite{DCN}, factorization machine~\cite{DeepFM}, or inner product~\cite{IPNN} can be chosen. Similarly, deep components may also vary from MLP layer~\cite{FNN} to self-attention module~\cite{Autoint}. Below, we separately discuss the two parameters of fusion learning: fusion connection $\mathbf{c}$ and fusion operation $\mathbf{o}$.

\subsubsection{Fusion Connection} 
Fusion connection $\mathbf{c}$ determines the connectivity between different components. Such connectivity determines what information is fed into the current component. There are only two potential states of connectivity between two components, i.e., \texttt{CONNECTED} or \texttt{DISCONNECTED}. Hence, we adopt a connectivity function $\mathbf{c}(\cdot)$ to formulate the fusion connection from component $\mathcal{G}_i$ to component $\mathcal{G}_j$ as 
\begin{equation}
\label{eq:connectivity}
\mathbf{c} \left( \mathcal{G}_i, \mathcal{G}_j \right) =
\begin{cases}
    1, & \text{if  } \texttt{CONNECTED} \\
    0, & \text{if  } \texttt{DISCONNECTED} 
\end{cases}
\end{equation}

In the fusion learning, each component has its level, which constrains the direction of the connection. The current component only takes the outputs from lower-level components as the input for the fusion module. Such design is introduced to avoid cycles~\cite{AutoCTR,AssembleNet++}. Such a constraint can be formulated on each $\left( \mathcal{G}_i, \mathcal{G}_j\right)$ pairs:
\begin{equation}
\label{eq:constraint}
    \mathbf{c} \left( \mathcal{G}_i, \mathcal{G}_j \right) = 1 \rightarrow L(i) < L(j), \forall \left( \mathcal{G}_i, \mathcal{G}_j\right)
\end{equation}
where $L(i)$ is a non-negative integer denotes the level of component $\mathcal{G}_i$. It increases monotonically as the component gets deeper.

\subsubsection{Fusion Operation}
After determining the fusion connection, a fusion operation $\mathop{o}$ needs to be selected for each component. The fusion operation aggregates the output representations from the connected components. It outputs a fused representation, which serves as the input for the succeeding component $\mathcal{G}_j$, as shown in Figure \ref{fig:framework}. This process can be written as:
\begin{equation}
    \mathbf{\hat{e}}_j = o_j (\{ \mathbf{c} \left( \mathcal{G}_i, \mathcal{G}_j \right) \cdot \mathbf{e}_i \}).
\end{equation}
Here $\mathbf{\hat{e}}_j$ refers to the input for component $\mathcal{G}_j$ and $\mathbf{e}_i$ denotes the output for preceding component $\mathcal{G}_i$. Note that the fusion operations are usually selected from a set of candidates $\mathcal{O}$. It may vary from simple operations such as \texttt{ADD} or \texttt{CONCAT} to complex operations like Multi-Head Bilinear Fusion Ops~\cite{FinalMLP}. Given the one-to-one mapping relationship between fusion operation $\mathop{o}_j$ and component $\mathcal{G}_j$, we have $|\mathbf{o}| = |\mathcal{G}|$.

\subsubsection{Fusion Learning in CTR}
After formulating fusion connection learning and fusion operation selection, the prediction in Eq.~\ref{eq:output_naive} can be reformulated as:
\begin{equation}
\label{eq:output_fusion}
     \hat{y} = \mathcal{F}(\mathcal{E}(x), \Theta) = \mathcal{P}(\mathcal{G}|\mathbf{c}, \mathbf{o})(\mathcal{E}(x), \Theta).
\end{equation}
Consequently, the training objective in Eq.~\ref{eq:loss_naive} can be rewritten as:
\begin{equation}
\label{eq:loss_fusion}
\begin{aligned}
\argmin_{\Theta, \ \mathbf{c} \in \{0, 1\}, \ \mathbf{o} \in \mathcal{O}} \mathcal{L}_{\mathcal{D}}(\Theta, \mathbf{c}, \mathbf{o})  \\
\end{aligned}
\end{equation}

\section{OptFusion}
\label{sec:method}
In this section, we detail the OptFusion framework under the fusion learning setting. We first describe the search space of the framework in Section \ref{sec:method_framework}. Then, we introduce fusion connection learning in Section~\ref{sec:connection_search} and fusion operation selection in Section~\ref{sec:fusion_search}. 
Finally, we elaborate on the details of the one-shot learning algorithm, which jointly conducts fusion connection learning and fusion operation selection for OptFusion,
in Section~\ref{sec:learning_algo}.

\subsection{Search Space}
\label{sec:method_framework}
In this section, we detail the search space of the framework in the fusion learning setting.
The OptFusion framework consists of one embedding component $\mathcal{E}(\cdot)$, $n$ shallow components, $n$ deep components, and one output component $\mathcal{H}(\cdot)$. The number of all components in the search space is $2n+2$. 
The default configuration of the OptFusion framework with $n=3$ is illustrated in Figure \ref{fig:framework} (b).
The setting of $n$ is discussed in Section~\ref{sec:number_n}.

\textit{Candidates of Fusion operation.} Shallow and deep components may receive multiple inputs from lower-level components. Thus, a fusion operation is needed to fuse these inputs. The commonly-used fusion operations include \texttt{ADD}, \texttt{PROD}, \texttt{CONCAT} and \texttt{ATT}, which represent element-wise addition, Hadamard product, concatenation, and attention, respectively. Details are shown in Table \ref{tab:operations}. Each component can select one of them to fuse information from lower-level components. Alternatively, each component can also output a weighted sum. Depending on this, we propose two variants of OptFusion, namely \textit{Hard} and \textit{Soft}.

\begin{table}[!htbp]
\centering
\vspace{-10pt}
\caption{A summary of different fusion operations.}
\vspace{-10pt}
\resizebox{.45\textwidth}{!}{
\begin{tabular}{c|c|c}
     \hline
     Operation & Description & Formulation \\
     \hline
     \texttt{ADD} & Element-wise addition & $\mathbf{x}' = \mathbf{x}_{i} + \mathbf{x}_{j}$ \\
     \hline
     \texttt{PROD} & Hadamard product & $\mathbf{x}' = \mathbf{x}_{i} \odot \mathbf{x}_{j}$ \\
     \hline
     \texttt{CONCAT} & Concatenation & $\mathbf{x}' = \mathbf{W} [\mathbf{x}_{i} || \mathbf{x}_{j}]$ \\
     \hline
     \multirow{2.6}{*}{\texttt{ATT}} & \multirow{2.6}{*}{Attention mechanism} & $\mathbf{x}' = a_i \cdot \mathbf{x}_{i} + a_j \cdot \mathbf{x}_{j}$ \\
     ~ & ~ & $a_i = \frac{{\rm exp} (\mathbf{w}_{2}^{\mathsf{T}} {\rm ReLU}(\mathbf{W}_{1} \mathbf{x}_{i} + \mathbf{b}_{1}))}{\sum_{j} {\rm exp} (\mathbf{w}_{2}^{\mathsf{T}} {\rm ReLU}(\mathbf{W}_{1} \mathbf{x}_{j} + \mathbf{b}_{1}))}$ \\
     \hline
\end{tabular}
\label{tab:operations}}
\begin{tablenotes}\footnotesize
\item Take an example to illustrate the fusion operation, two vectors $\mathbf{x}_{i} \in \mathbb{R}^{d}$ and $\mathbf{x}_{j} \in \mathbb{R}^{d}$ are used as inputs, and $\mathbf{x}' \in \mathbb{R}^{d}$ is the output of fusion. In \texttt{PROD}, $\odot$ denotes the Hadamard product operation. In \texttt{CONCAT}, $||$ denotes the concatenation operation, and $\mathbf{W} \in \mathbb{R}^{d \times 2d}$ is the trainable weight parameter. In \texttt{ATT}, $a_i$ and $a_j$ are attention coefficients, $\mathbf{W}_{1} \in \mathbb{R}^{d \times d}$, $\mathbf{w}_{2} \in \mathbb{R}^{d}$ and $\mathbf{b}_{1} \in \mathbb{R}^{d}$ are the trainable weight and bias parameters.
\end{tablenotes}
\vspace{-10pt}
\end{table}

\textit{Search space analysis.} 
In this subsection, we intuitively illustrate the difficulty in selecting suitable fusions.
Given that OptFusion aims to search for both fusion connections and operations, we need to jointly consider their possibility. The number of possible connections is determined by the number of components $2n+2$. The number of all valid connections equals to 
$2 \times (1+ 3+ \cdots +2n-1)+2(n+1) = 2n^2+2n+1$. For each valid connection, the number of choices for its connection state is 2. Thus, the size of the search space for connections is 
$2^{2n^2+2n+1}$. 
For each component, suppose the number of possible choices for its fusion operation is $k$. Thus, the search space size for the fusion operation is $k^{2n+1}$. The number of possible fusions equals $2^{2n^2+2n+1} \times k^{2n+1} = \mathbf{O}(2^{2n} \times k^{n})$. Directly selecting over such a large space is almost impossible. Hence, we separately discuss the connection learning and operation selection in the following two sections.

\subsection{Fusion Connection Learning}
\label{sec:connection_search}

A critical issue for fusion connection learning lies in the discrete selection space. 
Searching within a discrete candidate set of connections (i.e., $\mathcal{C}=\left\{\texttt{CONNECTED}, \texttt{DISCONNECTED}\right\}$) is non-differentiable, which makes the architecture untrainable. 
To solve this problem, we relax the discrete search space to be continuous by learning the relative importance (i.e., probability) of each connection and introduce the architecture parameters $\boldsymbol{\alpha} \in \mathbb{R}^{(2n+2)^2}$ to parameterize the connectivity function $\mathbf{c}(\cdot)$ so that the fusion connection becomes learnable. With $\boldsymbol{\alpha}_{ij}$ representing the connectivity $\mathbf{c}(\mathcal{G}_i, \mathcal{G}_j)$ from component $\mathcal{G}_i$ to $\mathcal{G}_j$, Eq.~\ref{eq:connectivity} can be rewritten as follows
\begin{equation}
\boldsymbol{\alpha}_{ij} 
\begin{cases}
    >   0, & \text{if  } \texttt{CONNECTED} \\
    \le 0, & \text{if  } \texttt{DISCONNECTED} 
\end{cases}
, \ 1 \le i,j \le 2n+2.
\end{equation}
To satisfy the level constraint in Eq.~\ref{eq:constraint}, $\boldsymbol{\alpha}$ is constrained as:
\begin{equation}
    \boldsymbol{\alpha}_{ij} \ge 0 \rightarrow L(i) < L(j), \ \forall 1 \le i, j \le 2n+2.
\end{equation}


To enable end-to-end training and get meaningful gradients for $\boldsymbol{\alpha}$, we adopt the straight-through estimator (STE) function~\cite{STE}. The STE can be formulated as a customized function S(·), with its forward pass as a unit step function $S(x) = 0, x \leq 0$ and $S(x) = 1, x > 0$.
$S(x)$'s backward pass equals to $\frac{d}{d x} \mathcal{S}(x)=1$, meaning that it will directly pass the gradient backward. Therefore, we can mimic a discrete selection while providing valid gradient information for connection parameters $\boldsymbol{\alpha}$, making the whole process trainable. Hence, the final output in Eq.~\ref{eq:output_fusion} can be rewritten as:
\begin{equation}
    \hat{y} = \mathcal{P}(\mathcal{G}|\boldsymbol{\alpha}, \mathbf{o})(\mathcal{E}(x), \Theta).
    \label{eq:output_alpha}
\end{equation}

\subsection{Fusion Operation Selection}
\label{sec:fusion_search}
For shallow or deep components, they may receive multiple connections from lower-level components. One operation needs to be selected from the set of fusion operations to fuse the information received from lower-level components. Specifically, we define the operation candidates as $\mathcal{O}$ = $\{\texttt{ADD}$, $\texttt{PROD}$, $\texttt{CONCAT}$, $\texttt{ATT}\}$ and $|\mathcal{O}| = k$. Similar to connection learning, we also relax the discrete search space of fusion operations to be continuous by learning the relative importance of each operation and introduce the architecture parameters $\boldsymbol{\beta} \in \mathbb{R}^{k \times (2n+2)}$ to represent the operation selection. 

Given a component $j$, we assign an architecture parameter $\beta_{j}^{o}$ to an operation $o \in \mathcal{O}$, the importance of operation $o$ is computed as a \textit{softmax} of all candidate operations $o' \in \mathcal{O}$: 
\begin{equation}
    p_{j}^{o} = {\rm exp} \left(\beta_{j}^{o} \right) / \sum_{o' \in \mathcal{O}} {\rm exp} \left(\beta_{j}^{o'} \right),
    \label{eq:fusion_operation}
\end{equation}
where $p_{j}^{o}$ is the importance of operation $o$. 
During the selection stage, the input of component $j$ equals a weighted summation over all candidate operations:
\begin{equation}
    \mathbf{\hat{e}}_j = \sum_{\mathop{o} \in \mathcal{O}} p_j^o \cdot \mathop{o} \left( \{ \boldsymbol{\alpha}_{ij} \cdot \mathbf{e}_i \} \right),
    \label{eq:weighted_summation}
\end{equation}
where $\mathbf{e}_{i}$ is the output of component $i$. $\mathop{o}(\cdot)$ fuses all the inputs by using operation $\mathop{o}$. Finally, by parameterizing the selection of fusion operation $\mathop{o} \in \mathcal{O}$ with architecture parameters $\boldsymbol{\beta}$, Eq.~\ref{eq:output_alpha} can be rewritten as follows:
\begin{equation}
    \hat{y} = \mathcal{P}(\mathcal{G}|\boldsymbol{\alpha}, \boldsymbol{\beta})(\mathcal{E}(x), \Theta).
    \label{eq:output_final}
\end{equation}

\subsection{One-shot Learning Algorithm}
\label{sec:learning_algo}

After obtaining Eq.~\ref{eq:output_final}, which parameterizes the fusion learning via architecture parameter $\{\boldsymbol{\alpha}, \boldsymbol{\beta}\}$, we need to rewrite the original learning goal in Eq.~\ref{eq:loss_fusion} to incorporate the fusion learning process. The optimization process can be rewritten as:
\begin{equation} \label{eq:loss_search}
    \mathop{\min}\limits_{\boldsymbol{\Theta}, \{\boldsymbol{\alpha}, \boldsymbol{\beta}\}} \mathcal{L}_{\mathcal{D}}\left(\boldsymbol{\Theta}, \{\boldsymbol{\alpha}, \boldsymbol{\beta}\}\right)
\end{equation} 

We can observe that the parameters that need to be optimized include the following three categories:
\begin{itemize}[topsep=0pt,noitemsep,nolistsep,leftmargin=*]
    \item $\boldsymbol{\Theta}$, the model parameters, including the parameters in all components and the transformation matrices between components. 
    \item $\{ \boldsymbol{\alpha}, \boldsymbol{\beta}\}$, the architecture parameters of fusion design, representation connection learning and operation selection, respectively.
\end{itemize}

Firstly, to disentangle the model and architecture parameters, we follow the paradigm from NAS~\cite{DARTS,AutoCTR} by first learning the architecture parameters $\{\boldsymbol{\alpha}, \boldsymbol{\beta}\}$ and conduct retraining to get model parameter $\boldsymbol{\Theta}$ given the architecture parameters.


Secondly, the entanglement between fusion connection and fusion operation remains a challenge. 
An intuitive approach is determining fusion connection $\boldsymbol{\alpha}$ and fusion operation $\boldsymbol{\beta}$ sequentially. However, such a design omits the mutual influence between connection and operation, as selecting inappropriate operations would decrease the likelihood of connections. 

Finally, we formulate the one-shot learning algorithm, which jointly and simultaneously conducts connection learning and operation selections given their entanglement. The one-shot learning algorithm consists of two stages: \textit{selection stage} and \textit{re-train stage}. In the selection stage, connection learning and operation selection are achieved by simultaneously optimizing the architecture parameters of connections $\boldsymbol{\alpha}$ and fusion operations $\boldsymbol{\beta}$. In the re-train stage, the architecture parameters $\boldsymbol{\alpha}$ and $\boldsymbol{\beta}$ are fixed, and the model parameter $\Theta$ is re-trainted.

\subsubsection{Selection Stage}
The goal in the selection stage is to jointly learn $\boldsymbol{\alpha}$ and $\boldsymbol{\beta}$ given their mutual information. This allows the model to explore different connections and fusion operations during the selection process. The optimization can be formulated as below:
\begin{equation}
    \boldsymbol{\alpha}^*, \boldsymbol{\beta}^* =  \argmin \limits_{\boldsymbol{\Theta}, \{\boldsymbol{\alpha}, \boldsymbol{\beta}\}} \mathcal{L}_{\mathcal{D}}\left(\boldsymbol{\Theta}, \{\boldsymbol{\alpha}, \boldsymbol{\beta}\}\right)
    \label{eq:selection_stage}
\end{equation}

\subsubsection{Re-train Stage}
In the retraining stage, the parameters need to be optimized only to include the model parameters $\boldsymbol{\Theta}$. We keep the selected architecture parameters $\boldsymbol{\alpha}^{*}$ and $\boldsymbol{\beta}^{*}$ fixed, re-train the model parameters $\boldsymbol{\Theta}$ to obtain the final model. Following previous works~\cite{DARTS}, the selected connection tensor is determined as 
$\alpha^*=\mathds{1}_{\alpha_{i,k}>0}$ during the re-training stage.
With the different ways to conduct fusion operations based on the score, we propose two variants of the proposed model, i.e., OptFusion-Soft and OptFusion-Hard, which refer to soft selection and hard selection of the fusion operations, respectively.

\textit{OptFusion-Soft}. In OptFusion-Soft, fusion operations are performed in a soft manner, i.e., combining the weighted summation over all candidate operations according to Equation~\ref{eq:weighted_summation}. The final architecture parameter for fusion can be formulated as $\beta^* = \beta$.

\textit{OptFusion-Hard}. The final fusion operation type is selected with the largest weight based on the learned fusion parameters. This is formalized as:
$\beta^{o*}_k = 1$ if $o = \underset{o \in \mathcal{O}}{\arg \max } \beta_{k}^{o}$ and $\beta^{o*}_k = 0$ otherwise.

After obtaining the architecture parameters $\alpha^*$ and $\beta^*$ for fusion connection and operation.
The model parameters are then re-trained with fixed architecture parameters.
\begin{equation}
    \boldsymbol{\Theta}^* = \argmin\limits_{\boldsymbol{\Theta}} \mathcal{L}_{\mathcal{D}}\left(\boldsymbol{\Theta}, \boldsymbol{\alpha}^{*}, \boldsymbol{\beta}^{*}\right)
    \label{eq:retrain_stage}
\end{equation}

This one-shot selection algorithm allows OptFusion to efficiently explore different architectures during the selection stage and then fine-tune the discovered architecture in the re-train stage. Finally, the pseudo-code of the learning algorithm for OptFusion is summarized in Algorithm~\ref{alg:OptFusion}.

\begin{algorithm}[!htbp]
	\caption{The OptFusion Algorithm} 
    \label{alg:OptFusion}
	\begin{algorithmic}[1]
		\Require Training dataset $\mathcal{D}$ consisting original features $x$ and ground-truth labels $y$
        \Ensure Learned architecture parameter $ \boldsymbol{\alpha}^{*}, \boldsymbol{\beta}^{*}$, model parameters $\boldsymbol{\Theta}^*$
        \State \textbf{Selection Stage}
        \State t=0
        \While {t < T}
            \State t = t + 1
            \While{not converged}
                \State Sample a mini-batch $\mathcal{B}$ from the training dataset $\mathcal{D}$
                \State Update the model parameters $\boldsymbol{\Theta}$, connection parameters $\boldsymbol{\alpha}$ and operation parameters $\boldsymbol{\beta}$ by Eq.~\ref{eq:selection_stage}
            \EndWhile
        \EndWhile 

        \State \textbf{Re-train Stage}

        \State Retrain $\boldsymbol{\Theta}$ given $\boldsymbol{\alpha}^{*}, \boldsymbol{\beta}^{*}$ by Eq.~\ref{eq:retrain_stage}
	\end{algorithmic}
\end{algorithm}

\section{Experiments}
\label{sec:experiment}
In this section, to comprehensively validate OptFusion, we design and conduct various experiments over three large-scale datasets, aiming to answer the following research questions:
\begin{itemize}[leftmargin=*]
    \item \textbf{RQ1}: Could OptFusion achieve superior performance compared with mainstream deep CTR prediction models?
    \item \textbf{RQ2}: How efficient is OptFusion compared with mainstream deep CTR prediction models?
    \item \textbf{RQ3}: How does the selection of fusion operation influence the performance?
    \item \textbf{RQ4}: How effective is the one-shot selection algorithm?
    \item \textbf{RQ5}: How compatible is OptFusion with existing components?
    \item \textbf{RQ6}: How does the number of components affect performance?
    \item \textbf{RQ7}: Does OptFusion select the suitable fusion?
\end{itemize}

\subsection{Experimental Setting}
\subsubsection{Datasets}
To demonstrate the effectiveness of OptFusion, we evaluate our model on three real-world datasets, the statistics of datasets are described in Table~\ref{tab:datasets}.
\begin{table}[!htbp]
\caption{Statistics of datasets.}
\vspace{-10pt}
\centering
\begin{tabular}{c|cccccc}
    \hline
    Dataset & \#samples & \#Fields & \#Values & pos ratio \\
    \hline
    Criteo      & $4.6 \times 10^7$ & 39 & $6.8 \times 10^6$ & 0.2562 \\
    Avazu       & $4.0 \times 10^7$ & 24 & $4.4 \times 10^6$ & 0.1698 \\
    KDD12       & $1.5 \times 10^8$ & 11 & $6.0 \times 10^6$ & 0.0445 \\
    \hline
\end{tabular}
\begin{tablenotes}
\footnotesize
\item[1] Note: \textit{\#samples} refers to the total samples in the dataset, \textit{\#field} refers to the number of feature fields for original features, \textit{\#value} refers to the number of feature values for original features, \textit{pos ratio} refers to the positive ratio. 
\end{tablenotes}
\label{tab:datasets}
\vspace{-5pt}
\end{table}

\textbf{Criteo}\footnote{\url{https://www.kaggle.com/c/criteo-display-ad-challenge}} dataset consists of ad click data over a week. It consists of 26 categorical feature fields and 13 numerical feature fields. Following the winner solution of the Criteo Advertising Challenge~\cite{criteo-display-ad-challenge}, we discretize each numeric value $x$ to $\lfloor\log^2(x)\rfloor$, if $x>2$; $x=1$ otherwise. We replace infrequent categorical features with a default "OOV" (i.e. out-of-vocabulary) token, with \textit{min\_count}=2.

\textbf{Avazu}\footnote{\url{http://www.kaggle.com/c/avazu-ctr-prediction}} dataset contains 10 days of click logs. It has 24 fields with categorical features. We remove the \textit{instance\_id} field and transform the \textit{timestamp} field into three new fields: \textit{hour}, \textit{weekday} and \textit{is\_weekend}. We replace infrequent categorical features with the "OOV" token, with \textit{min\_count}=2.

\textbf{KDD12}\footnote{\url{http://www.kddcup2012.org/c/kddcup2012-track2/data}} dataset contains training instances derived from search session logs. It has 11 categorical fields, and the click field is the number of times the user clicks the ad. We replace infrequent features with an "OOV" token, with \textit{min\_count}=2.

\subsubsection{Metrics}
To evaluate the performance of CTR Prediction, we adopt the most commonly-used evaluation metrics~\cite{DeepFM}, i.e., \textbf{AUC} (Area Under ROC) and \textbf{LogLoss} (cross-entropy). Note that $\mathbf{0.1 \%}$ improvement in AUC is considered significant~\cite{IPNN, DeepFM, IncMSR}.

\subsubsection{Baselines}
To demonstrate the effectiveness of OptFusion, we compare the performance with four categories of deep CTR prediction models, including (i) stacked models: FNN~\cite{FNN} and PNN~\cite{IPNN}, DCNv2s~\cite{DCNv2}; (ii) parallel models: DeepFM~\cite{DeepFM}, DCN~\cite{DCN}, xDeepFM~\cite{xDeepFM}, DCNv2p~\cite{DCNv2}\footnote{DCNv2s refers to a stacked design while DCNv2p involves a parallel design in the original paper~\cite{DCNv2}.}; (iii) models with expert design on fusion: EDCN~\cite{EDCN}; (iv) NAS models: AutoCTR~\cite{AutoCTR}, NASRec~\cite{NASRec}.

\subsubsection{Implementation Details}
We use Adam~\cite{Adam} as the optimizer for all models and set the embedding size as 40 for the Criteo and Avazu datasets and 16 for the KDD12 dataset. The batch size is fixed at 4096. Following~\cite{EDCN}, we employ a three-layer MLP with the number of neurons equal to $dim_{emb} \times num_{field}$. We also incorporate an auxiliary shallow block $S0$ within the candidate set of connections to mimic the stacked structure.
We search the optimal learning rate from \{3e-3, 1e-3, 3e-4, 1e-4, 3e-5, 1e-5\} and $L_2$ regularization from \{3e-6, 1e-6, 3e-7, 1e-9, 0\}. For OptFusion, during the re-training phase, we reuse the optimal learning rate and $L_2$ regularization obtained in the initial training. $\alpha$ in connection search is initialized as 0.5 to ensure equal weight for each block at the start. Hyperparameters used in the experiments are reported in \footnote{\url{https://github.com/kexin-kxzhang/OptFusion/hyperparam.md}}, and the implementation of our algorithm is available here\footnote{\url{https://github.com/kexin-kxzhang/OptFusion}}. For NAS models, we reuse the official implementation\footnote{\url{https://github.com/facebookresearch/nasrec}} with the same embedding size and batch size as ours.



\subsection{Overall Performance (RQ1)}
The overall performance of our OptFusion and other deep CTR prediction models on three datasets are reported in Table~\ref{tab:overall}. We summarize the observations as follows:

\begin{table}[!htbp]
\centering
\caption{The overall performance comparison.}
\vspace{-10pt}
\resizebox{.49\textwidth}{!}{
\begin{tabular}{c|cc|cc|cc}
\hline
\multirow{2}{*}{Method} & \multicolumn{2}{c|}{Criteo} & \multicolumn{2}{c|}{Avazu} & \multicolumn{2}{c}{KDD12} \\ 
\cline{2-7} 
& AUC & Logloss & AUC & Logloss & AUC & Logloss \\ 
\hline
FNN & 0.8037 & 0.4473 & 0.7860 & 0.3766 & 0.7978 & 0.1534 \\
PNN & 0.8048 & 0.4463 & 0.7886 & 0.3752 & 0.8011 & 0.1527 \\
DCNv2s & 0.8088 & 0.4427 & 0.7877 & 0.3753 & 0.8030 & 0.1536 \\
\hline
DeepFM & 0.8038 & 0.4486 & 0.7856 & 0.3806 & 0.7963 & 0.1532 \\
DCN & 0.8063 & 0.4450 & 0.7875 & 0.3779 & 0.7968 & 0.1537 \\
xDeepFM & 0.8067 & 0.4453 & 0.7860 & 0.3773 & 0.7966 & 0.1542 \\
DCNv2p & 0.8085 & 0.4451 & 0.7894 & 0.3759 & 0.8012 & 0.1531 \\
\hline
EDCN & \underline{0.8102} & \underline{0.4419} & \underline{0.7917} & \underline{0.3727} & \underline{0.8122} & \underline{0.1498} \\
\hline
AutoCTR & 0.8082 & 0.4436 & 0.7883 & 0.3761 & 0.7949 & 0.1533 \\
NASRec & 0.8090 & 0.4435 & 0.7893 & 0.3752 & 0.7958 & 0.1530 \\
\hline
OptFu.-H & 0.8108* & 0.4413* & 0.7935* & 0.3717* & 0.8129 & 0.1496 \\
OptFu.-S & \textbf{0.8113*} & \textbf{0.4408*} & \textbf{0.7938*} & \textbf{0.3715*} & \textbf{0.8158*} & \textbf{0.1489*} \\
\hline
Impr & 0.0011 & 0.0023 & 0.0021 & 0.0012 & 0.0036 & 0.0009 \\
\hline
\end{tabular}
}
\begin{tablenotes}\footnotesize
    \item[*] Here $*$ denotes statistically significant improvement (measured by a two-sided t-test with $p$-value $<$ 0.05) over the best baseline. Bold scores are the best performance, and underlined scores are the best baseline performance. OptFu.-H and OptFu.-S stand for OptFusion-Hard and OptFusion-Soft, respectively.
\end{tablenotes}
\vspace{-5pt}
\label{tab:overall}
\end{table}

First, OptFusion, both soft and hard, outperforms all the SOTA baselines over three datasets in terms of both AUC and Logloss by a significant margin. This demonstrates that OptFusion can effectively find a suitable fusion connection and operation. It also echoes our intuition: fusion learning is an important but overlooked aspect of feature interaction modeling.
Specifically, OptFusion improves AUC over the best baseline by 0.0011, 0.0021, and 0.0036 on three datasets, respectively.

Second, OptFusion, with a smaller search space, outperforms NASRec and AutoCTR, which contain a larger search space. Such an observation demonstrated that OptFusion could better exploit and explore the fusion search space, while NASRec and AutoCTR's huge search space could potentially lead to sub-optimal results.

Third, EDCN, which aims to fuse explicit and implicit information densely, constantly performs as the best baseline over all three datasets. This proves that the naive fusion design is an obstacle towards accurate prediction, echoing previous observations~\cite{EDCN}.

Finally, the performance of naive fusions varies across datasets. 
For example, on the Avazu datasets, DCNv2p, a parallel model, exhibits superior performance. On the Criteo and KDD12 datasets, DCNv2s, a stack model, outperforms other baselines. This interesting observation further reveals the limitation of naive fusion design, which we will discuss in Section \ref{sec:exp_case_study}.

\subsection{Efficiency Analysis (RQ2)}
In addition to model effectiveness, training, and inference efficiency are crucial considerations when deploying CTR prediction models in practice. In this section, we investigate the time complexity of OptFusion. Due to the expansive search space of AutoCTR and NASRec, training efficiency experiments are conducted on an NVIDIA A40 GPU with 48G memory, while inference efficiency experiments are conducted on an NVIDIA RTX 4090 GPU with 24G memory.

\begin{figure*}[!htbp]
\centering
\includegraphics[width=0.95\textwidth]{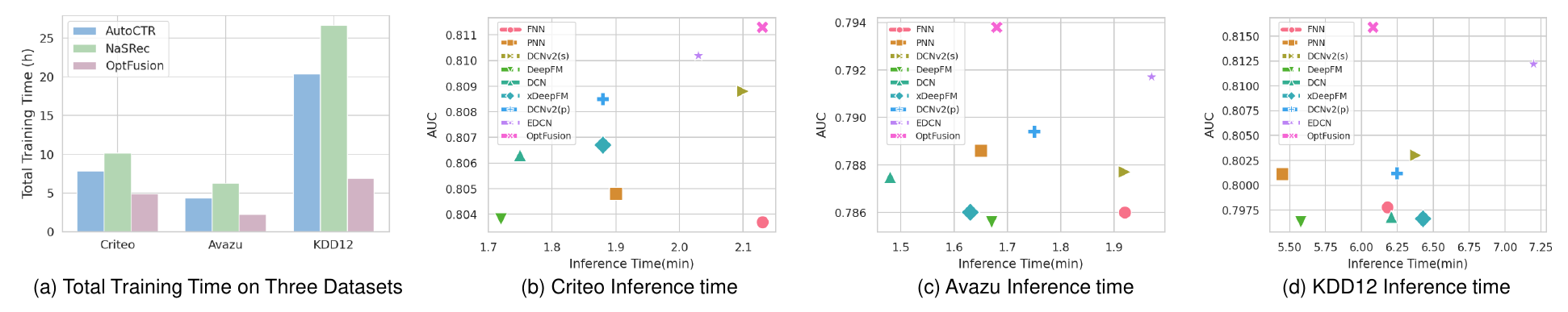}
\vspace{-10pt}
\caption{Efficiency analysis across three datasets. The figure includes (a) total training time on three datasets and (b--d) inference efficiency analysis for the Criteo, Avazu, and KDD12 datasets.}
\Description[Efficiency Analysis]{Combined analysis of training and inference efficiency on three datasets.}
\vspace{-5pt}
\label{fig:efficiency-analysis}
\end{figure*}

We illustrate the total training time of NAS models trained on all three datasets in Figure~\ref{fig:efficiency-analysis} (a). Here, the total training time encompasses both the search and re-train stages. We observe that OptFusion achieves the shortest total training time compared to other NAS models. This is attributed to the narrowed search space for OptFusion and the adoption of a one-shot learning algorithm. 

As depicted in Figure \ref{fig:efficiency-analysis} (b-d), we plot the Inference Time-AUC curve of mainstream deep CTR models trained on three datasets, indicating the relationship between time complexity and model performance. Compared with models such as DCN, DeepFM, and PNN, which achieve the least inference time, both EDCN and OptFusion take fusion design into consideration, and they tend to achieve the highest AUC. This can be attributed to the trade-off between inference time and model performance.

\subsection{Ablation Study}
\subsubsection{Fusion Operation (RQ3)}

In this section, we aim to investigate how the fusion operation influences the performance of deep CTR models. We reuse the searched fusion connection and replace the searched fusion operation with four identical fusion operations: ADD, PROD, CONCAT, and ATT, as introduced in Table ~\ref{tab:operations}. These models corresponding to four fusion operations are retrained from scratch. Results of OptFusion-soft and OptFusion-hard selection, which adopt different operations for different blocks, in Table \ref{tab:overall} are also listed for easy comparison. The results are shown in Table ~\ref{tab:ablation_fusion}.

We can easily observe that both Soft and Hard methods exhibit significantly superior performance compared to models with fixed fusion operations. This verifies the effectiveness of our fusion operation search instead of using a fixed fusion operation. 

In addition, ADD and PROD operations outperform the others. This may be attributed to the fact that element-wise addition and Hadamard products are parameter-free operations, making them easier to train steadily compared to methods incorporating parameters like concatenation and attention pooling.

\begin{table}[!htbp]
\centering
\caption{Ablation study on fusion operation.}
\vspace{-10pt}
\label{tab:ablation_fusion}
\resizebox{.48\textwidth}{!}{
\begin{tabular}{c|cc|cc|cc}
\hline
\multirow{2}{*}{Operation} & \multicolumn{2}{c|}{Criteo} & \multicolumn{2}{c|}{Avazu} & \multicolumn{2}{c}{KDD12} \\ 
\cline{2-7} 
            & AUC & Logloss & AUC & Logloss & AUC & Logloss \\ 
\hline
Add         & 0.8111 & 0.4422 & 0.7872 & 0.3970 & 0.7924 & 0.1585 \\
Product     & 0.8077 & 0.4443 & 0.7860 & 0.3784 & 0.7938 & 0.1584 \\
Concatenate & 0.8075 & 0.4445 & 0.7837 & 0.3814 & 0.7926 & 0.1546 \\
Attention   & 0.8073 & 0.4442 & 0.7843 & 0.3794 & 0.7883 & 0.1597 \\
\hline
Hard        & 0.8108 & 0.4413 & 0.7935 & 0.3717 & 0.8129 & 0.1496 \\
Soft        & \textbf{0.8113} & \textbf{0.4408} & \textbf{0.7938} & \textbf{0.3715} & \textbf{0.8158} & \textbf{0.1489} \\ 
\hline
\end{tabular}
}
\vspace{-10pt}
\end{table}

\subsubsection{Selection Algorithm (RQ4)}
In this section, we investigate the search algorithm design. We aim to compare the one-shot selection algorithm, which jointly selects both the connection and operation simultaneously, with a sequential selection algorithm, which sequentially selects the connection and operation. Experiments are conducted over Criteo and Avazu datasets, and the results are shown in Table ~\ref{tab:ablation_search}. 

\begin{table}[!htbp]
\centering
\caption{Ablation study on selection algorithm.}
\vspace{-10pt}
\label{tab:ablation_search}
\begin{tabular}{c|cc|cc}
\hline
\multirow{2}{*}{Methods} & \multicolumn{2}{c|}{Criteo} & \multicolumn{2}{c}{Avazu} \\ \cline{2-5} 
& AUC & Logloss & AUC & Logloss \\ \hline
One-shot & \textbf{0.8113} & \textbf{0.4408} & \textbf{0.7938} & \textbf{0.3715} \\
Sequential & 0.8109 & 0.4411 & 0.7934 & 0.3717 \\ \hline
\end{tabular}
\vspace{-5pt}
\end{table}

Results indicate that our one-shot selection algorithm performs better than the sequential selection algorithm. This gap is likely caused by the mutual influence between fusion connection and operation. Such an observation verifies the effectiveness of OptFusion and the one-shot selection algorithm.

\subsubsection{Shallow Component (RQ5)}
In this section, we conduct an ablation study on the compatibility of OptFusion over various explicit components. In the default setting, we adopt CrossNet~\cite{DCN} as the explicit component for OptFusion. We further replace CrossNet with CrossNetV2~\cite{DCNv2} and CIN~\cite{xDeepFM}, and construct its two variants: OptFusion-CrossNetV2 and OptFusion-CIN. The results are summarized in Table~\ref{tab:ablation_cross}. We additionally adopt the fusion connection and operation from EDCN as a comparison for all three explicit components, namely EDCN, EDCN-CrossNetV2, and EDCN-CIN.

\begin{table}[!htbp]
\centering
\caption{Ablation study on shallow components.}
\vspace{-10pt}
\label{tab:ablation_cross}
\begin{tabular}{c|cc|cc}
\hline
\multirow{2}{*}{Method} & \multicolumn{2}{c|}{Criteo} & \multicolumn{2}{c}{Avazu} \\ \cline{2-5} 
& AUC & Logloss & AUC & Logloss \\ \hline
DCN & 0.8063 & 0.4450 & 0.7895 & 0.3762 \\
EDCN & 0.8102 & 0.4419 & 0.7917 & 0.3727 \\
OptFu.-Soft & \textbf{0.8113} & \textbf{0.4408} & \textbf{0.7938} & \textbf{0.3715} \\
\hline
DCNv2 & 0.8085 & 0.4451 & 0.7903 & 0.3751 \\
EDCN-CrossNetV2 & 0.8091 & 0.4434 & 0.7923 & 0.3723 \\
OptFu.-CrossNetV2 & \textbf{0.8111} & \textbf{0.4408} & \textbf{0.7942} & \textbf{0.3717} \\
\hline
xDeepFM & 0.8067 & 0.4453 & 0.7860 & 0.3773 \\
EDCN-CIN & 0.8089 & 0.4426 & 0.7943  & 0.3712 \\ 
OptFu.-CIN & \textbf{0.8112} & \textbf{0.4408} & \textbf{0.7955} & \textbf{0.3703} \\ 
\hline
\end{tabular}
\begin{tablenotes}\footnotesize
    \item[*] OptFu. is an abbreviation for OptFusion.
\end{tablenotes}
\vspace{-10pt}
\end{table}

From the table, we can easily observe that OptFusion and its two variants achieve the best performance on both datasets. These results underscore the robustness and compatibility of OptFusion to different explicit components. Besides, EDCNs constantly rank the 2nd, outperforming the original models. This further indicates the importance of dense fusion in deep CTR models.

\subsubsection{Number of Components (RQ6)}
\label{sec:number_n}
This section evaluates the impact of varying the number of components ($n$) on OptFusion's performance. The default setting for $n$ is 3. To investigate the effect of different configurations, we also conduct experiments with $n=2$ and $n=4$. Table~\ref{tab:ablation_n} summarizes the results, including performance metrics and total training time (h) for each configuration across the Criteo and Avazu datasets.

Our observations indicate that increasing the number of components ($n$) results in a marginal improvement in performance metrics at the cost of total training time. For instance, with $n=4$, the training time is approximately 1.29 times longer than with $n=3$ and about 1.76 times longer than with $n=2$.
Based on these results, we choose $n=3$ as the default configuration for OptFusion, as it offers a balanced trade-off between performance and efficiency.
\begin{table}[!htbp]
\centering
\caption{Ablation study on the number of components.}
\vspace{-10pt}
\label{tab:ablation_n}
\resizebox{.49\textwidth}{!}{
\begin{tabular}{c|ccc|ccc}
\hline
\multirow{2}{*}{Number} & \multicolumn{3}{c|}{Criteo} & \multicolumn{3}{c}{Avazu} \\ \cline{2-7} 
& AUC & Logloss & Time (h) & AUC & Logloss & Time (h) \\ \hline
n=2 & 0.8112 & 0.4408 & 3.38h & 0.7937 & 0.3716 & 1.47h \\
n=3 & 0.8113 & 0.4408 & 4.70h & 0.7938 & 0.3715 & 1.98h \\
n=4 & 0.8115 & 0.4406 & 5.76h & 0.7939 & 0.3717 & 2.67h \\
\hline
\end{tabular}}
\vspace{-10pt}
\end{table}

\subsection{Case Study (RQ7)}
\label{sec:exp_case_study}
\begin{figure}[!htbp]
    \centering
    \includegraphics[width=0.44\textwidth]{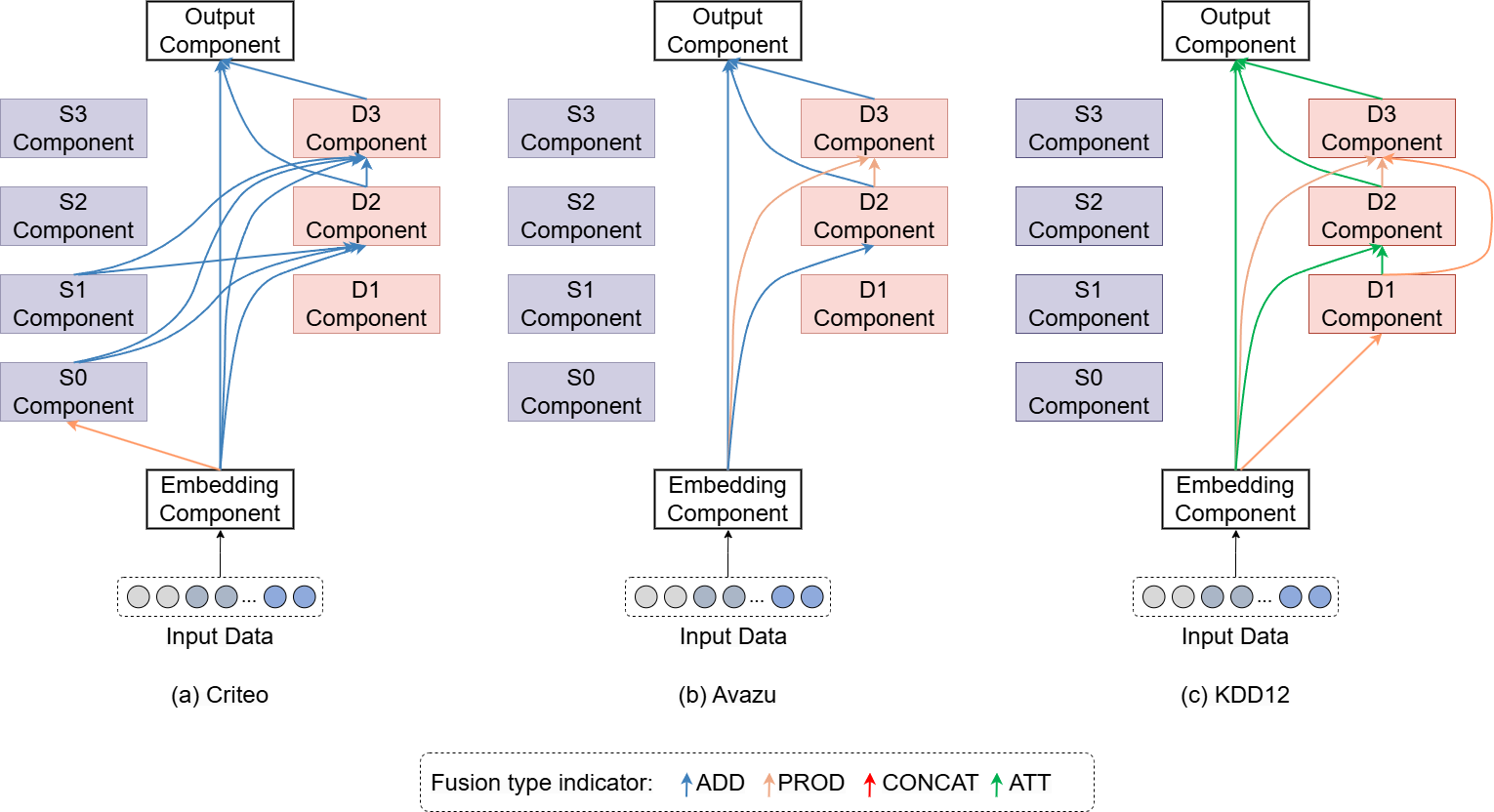}
    \vspace{-10pt}
    \caption{A case study of OptFusion on three datasets.}
    \vspace{-10pt}
    \Description[Case Study]{A case study of OptFusion on three datasets.}
    \label{fig:case}
\end{figure}
This section uses a case study to investigate the selected connection and operation obtained from OptFusion on three datasets. The selected results, including both connection and operation, are shown in Figure \ref{fig:case}. For better visualization, we only highlight the operation with the highest probability in OptFusion-soft, which is also the selected operation in OptFusion-hard. 

Based on the result, we can make the following observation:
First, the fusion searched on the Criteo dataset exhibits a preference for parallel structure, while on the Avazu and KDD12 datasets, the learned architectures indicate a prevalent inclination towards stacked structure.
Second, the output layer prefers to fuse connections from deep layers and the embedding layer. In addition, the embedding layer is connected to many other blocks. 
Third, the preference for fusion operation varies among datasets. ADD and PROD operations are the most common operations on the Criteo and Avazu datasets, while ATT and PROD are most frequently used on the KDD12 dataset. 
This case study provides valuable insights into the selected fusion by OptFusion, demonstrating the effectiveness and adaptability of OptFusion to different datasets. It may also provide design insights for future CTR models.

\section{Related Work}
\label{sec:related_work}

\subsection{Deep CTR Prediction Models}

Most CTR models adopt two naive design fusion designs~\cite{wang2022enhancing}, parallel and stacked. Models with parallel fusion~\cite{DCN,DCNv2,DeepFM,xDeepFM,Autoint} leverage shallow and deep components that explicitly and implicitly model feature interactions, respectively. Fusion operations mainly are addition~\cite{DeepFM} or concatenation~\cite{DCN,DCNv2}. Models with stacked fusion~\cite{IPNN,OENN,FGCNN,AFM,FiBiNet,DCNv2} tend to stack the shallow components before the deep components with concatenation being the common fusion operation~\cite{IPNN}. These models mainly advance CTR prediction by proposing various shallow components, such as inner product~\cite{IPNN}, factorization machine~\cite{DeepFM}, outer product~\cite{OENN}, convolutional operator~\cite{FGCNN}, Hadamard product~\cite{AFM} and different customized layers~\cite{DCN,DCNv2,FiBiNet,xDeepFM} or deep components, such as MLP~\cite{FNN} and Self-attention layer~\cite{Autoint}, to better model feature interactions.

Researchers also proposed methods with expert-designed fusion that are beyond parallel and stacked design. EDCN~\cite{EDCN} performs a dense fusion strategy and captures the layer-wise interactive signals between the deep and shallow components. FinalMLP~\cite{FinalMLP} proposes a Multi-Head Bilinear Fusion Ops as the fusion operation.
EulerNet~\cite{tian2023eulernet}, on the other hand, explores feature interaction learning using Euler’s formula, enabling adaptive and efficient fusion of feature interactions in CTR prediction models.
However, these proposed solutions~\cite{EDCN,FinalMLP} tend to consider the fusion learning problem under specific settings. 

OptFusion differs itself by automatically learning connections and selecting operations. 
Many of the aforementioned methods can be considered as specific instances of the OptFusion framework.

\subsection{Neural Architecture Search and its Applications in CTR Prediction}

With the advancement of neural architecture search (NAS)~\cite{DARTS,NAO,STE,FBNet,Gumble-softmax}, various methods have been proposed in CTR prediction~\cite{tang2023automl}, proving valuable for tasks such as determining appropriate embedding dimensions~\cite{joglekar2020neural}, conducting feature selection~\cite{OptFS,MultiFS}, discovering beneficial feature interactions~\cite{Autofis,OptFeature}, selecting integration function~\cite{Autofeature,OptInter}, optimizing hyperparameters~\cite{AutoOpt}, designing comprehensive architectures for feature interaction modeling~\cite{meng2021general},  or learning suitable embedding table~\cite{OptEmbed}. Various techniques such as evolutionary approach~\cite{real2017large}, gradient approach~\cite{DARTS}, or reinforcement learning-based methods~\cite{zoph2016neural} are introduced to obtain suitable search results.
Specifically, NAS techniques have also been adopted to search for suitable CTR model structures~\cite{AutoCTR,NASRec}.
Our work distinguishes itself from the existing research by addressing the challenge of fusion learning, a different problem in deep CTR models. The connection learning and operation selection among components are introduced as OptFusion's search space, enabling more efficient and effective model architectures for CTR prediction.

\section{Conclusion}
\label{sec:conclusion}
In this paper, we address the challenges of fusion learning in deep CTR prediction models and propose OptFusion, which automatically selects suitable fusion connections and fusion operations. 
OptFusion involves a one-shot learning algorithm designed to effectively conduct both tasks. The model is subsequently retrained with the learned architecture.
Extensive experiments on three large-scale datasets demonstrate the superior performance of OptFusion in terms of efficiency and effectiveness. Several ablation studies investigate the configuration of OptFusion in improving prediction performance. Additionally, a case study on the connection and fusion operations further validates the efficacy of our approach in learning suitable architectures.

\newpage
\section*{Ethical Consideration}
In conducting our research and proposing the OptFusion approach, we have adhered to ethical considerations to ensure the integrity and social responsibility of our work. Our research primarily focuses on advancing the technical aspects of recommendation systems. Our experiment is based on public benchmarks and does not involve direct interactions with human subjects or the collection of personal data. As such, potential ethical concerns related to informed consent, privacy, and data handling are minimized.

Our work aims to contribute to the field of recommendation systems by addressing technical challenges associated with automatically learning suitable fusions in click-through rate prediction models. Throughout our research process, we have followed established research ethics guidelines and practices to ensure the accuracy, transparency, and rigor of our methods and results. We have also taken care to properly attribute prior works and provide appropriate citations to relevant sources to maintain academic integrity.

We acknowledge that while our research primarily concerns technical advancements, the deployment and application of recommendation systems in real-world scenarios may raise broader ethical considerations related to user privacy, fairness, and potential algorithmic biases. We recognize their significance and encourage researchers and practitioners to approach the deployment of recommendation systems with careful consideration of these ethical implications.

In summary, our research on OptFusion has been conducted with a commitment to upholding ethical standards within the scope of our technical contributions.
\bibliographystyle{ACM-Reference-Format}
\balance
\bibliography{optfusion.bib}

\end{document}